\documentclass[aps,showpacs,preprint,preprintnumber,nofootinbib,amsmath,amssymb,ascmac,bm,12pt]{revtex4}

\usepackage{bm,amsmath,amssymb}

\usepackage[dvips]{graphicx}

\begin{document}

{~}

\vspace{1cm}

\title{
\Large 
Hawking radiation as tunneling \\ from squashed Kaluza-Klein black hole   
\vspace{0.5cm}
}

\author{
\large 
${}^{1}$Ken Matsuno\footnote{E-mail: matsuno@sci.osaka-cu.ac.jp} and  
${}^{2}$Koichiro Umetsu\footnote{E-mail: umetsu@cc.kyoto-su.ac.jp}
\vspace{0.5cm}
}	

\address{
${}^{1}$Department of Mathematics and Physics, Graduate School of Science, 
  Osaka City University, 
  3-3-138 Sugimoto, Sumiyoshi-ku, Osaka 558-8585, Japan
\\
${}^{2}$Maskawa Institute for Science and Culture, Kyoto Sangyo University, 
  Kyoto 603-8555, Japan
\vspace{1cm}
}

\begin{abstract}

We discuss Hawking radiation 
from a five-dimensional squashed Kaluza-Klein black hole on the basis of the tunneling mechanism. 
A simple manner, which was recently suggested by Umetsu, 
is possible to extend the original derivation by Parikh and Wilczek to various black holes. 
That is, we use the two-dimensional effective metric, 
which is obtained by the dimensional reduction near the horizon, as the background metric. 
By using same manner, we derive both the desired result of the Hawking temperature 
and the effect of the back reaction associated with the radiation
in the squashed Kaluza-Klein black hole background. 

\end{abstract}

\preprint{OCU-PHYS 344}  
\preprint{AP-GR 88} 
\preprint{MISC-2011-01} 


\date{\today}

\maketitle

\section{Introduction}

In 1975, Hawking 
showed that a black hole can radiate 
from the event horizon like a blackbody at the temperature 
$T = \kappa / (2 \pi)$, where $\kappa$ is the surface gravity of the black hole, 
by using the method of quantum field theory in curved spacetime \cite{Hawking:1974sw}. 
Thus Hawking radiation is one of interesting phenomena 
where both of general relativity and quantum theory play a role.

An elegant derivation of Hawking radiation on the basis of the tunneling method 
was proposed by Parikh and Wilczek \cite{Parikh:1999mf}.  
The derivation of Hawking radiation on the basis of the tunneling effect 
has been discussed in the literature 
\cite{SP,T01-1,T02,T02-1,T02-2,T03,T04,T05,T06,T07,T07-1,T07-2,T07-3,T07-4,T09,T09-1,T09-2,T10,
T10-1,T10-2,T11,T12,T13,T14,T15,T15-1,T16,T17,T18,T18-1,T18-2,T18-3,T18-4,T18-5,T18-6,T18-7,T18-8,
T19,T19-1,T19-2,T20,T21,T22,T23,T24,T25,T26,T27,T28,T29,
Wu:2007ty,Volovik:1999fc,
Vagenas:2000am,Vagenas:2001sm,Vagenas:2001rm,Vagenas:2002hs,Setare:2003vs,Setare:2004us,Medved:2005vw,
Banerjee:2009pf,Banerjee:2009sz,Roy:2009vy,Zhang:2010zzg,Ren:2010zzc,Majhi:2009xh,Banerjee:2010be,Banerjee:2010xi,
Yale:2008kx,Yale:2010qq,Yale:2010tn,
Zhang:2005wn,Zhang:2005gja,Jiang:2005ba,Chen:2007zzw}.    
The essential idea of the tunneling mechanism is that 
a particle-antiparticle pair is formed close to the horizon inside a black hole. 
We can divide the field associated with a particle into ingoing modes moving toward 
the center of the black hole and outgoing modes trying to move outside the horizon.   
Then the ingoing modes are trapped inside the horizon,  
while a part of the outgoing modes escapes outside the black hole by the quantum tunneling effect. 
If the particle which comes out to our universe has positive energy, 
such a particle (and also an antiparticle) can stably exist and we can regard the particle outside 
the horizon as the radiation from the black hole. 
Parikh and Wilczek calculated the WKB probability amplitude for the particle 
by taking into account classically forbidden paths. 
By comparing the tunneling probability of an outgoing particle 
with the Boltzmann factor in thermodynamics, 
they successfully derived the Hawking temperature. 
The derivation on the basis of the tunneling method   
has an important advantage that it is possible to evaluate the back reaction associated with the radiation.

Parikh and Wilczek obtained above results for 
four-dimensional Schwarzschild and Reissner-Nordstr\"om black holes. 
The extensions of their derivation to four-dimensional Kerr and Kerr-Newman black holes 
have been discussed in \cite{Zhang:2005wn,Zhang:2005gja,Jiang:2005ba,Chen:2007zzw}.  
The derivations for four-dimensional rotating black holes 
are mathematically complicated because of the effects of the rotation.  
Recently, 
one of the authors has suggested a new approach, 
which extends the method proposed by Parikh and Wilczek to four-dimensional rotating black holes,  
and showed that  
the derivation of the Hawking temperature is greatly simplified 
if one uses the technique of the dimensional reduction \cite{Umetsu:2010ts,Umetsu:2010kw}.
This technique has been used in the derivation of Hawking radiation from anomalies and    
discussed for various black holes   
(for example, see \cite{Iso:2006wa,Iso:2006ut,Iso:2006xj,Umetsu:2008cm,Lin:2010zza}).
According to the technique, 
it is shown that metrics of various black holes effectively become 
two-dimensional spherically symmetric metrics near the horizon.
We note that this technique of the dimensional reduction is valid only 
for the region very close to the horizon. 
Since the tunneling effect is a quantum one arising within the Planck length near the horizon region, 
the use of the technique of the dimensional reduction in the tunneling mechanism is justified.  
It is thus possible to naturally extend the method of Parikh and Wilczek to cases of various black holes
and we can derive both the Hawking temperatures and the effects of the back reactions 
associated with the radiations in various black hole backgrounds by using the new approach.
Then the derivations of the Hawking temperatures from four-dimensional rotating black holes   
become simple in comparison with previous ones \cite{Zhang:2005wn,Zhang:2005gja,Jiang:2005ba,Chen:2007zzw}.

A family of squashed Kaluza-Klein black holes is one of interesting solutions of higher-dimensional black holes. 
In the context of string theory and brane world scenario, 
investigations on higher-dimensional black hole solutions have attracted a lot of attention.  
From a realistic point of view,  
the extra dimensions need to be compactified to reconcile the higher-dimensional theory of gravity 
with our apparently four-dimensional world.  
We consider higher-dimensional black holes in the spacetime with compact extra dimensions, 
i.e., Kaluza-Klein black holes.  
In higher-dimensional spacetimes, 
even if we impose asymptotic flatness to the four-dimensional part of the spacetime, 
there are various possibilities of fiber bundle structures 
of the extra dimensions as the fiber over the four-dimensional base spacetime.  
The black hole solutions with nontrivial bundle structures 
have been studied by various authors. 
Dobiasch and Maison found the first five-dimensional vacuum Kaluza-Klein black hole solution
which asymptotes to a twisted $S^1$ fiber bundle 
over the four-dimensional Minkowski spacetime \cite{Dobiasch:1981vh}, 
and Gibbons and Wiltshire clarified the meaning of the metric \cite{Gibbons:1985ac}.  
Later, the solution was generalized to the charged case 
in five-dimensional Einstein-Maxwell theory \cite{Ishihara:2005dp}. 
It is shown that this family of black holes has squashed $S^3$ horizons, 
and that the metrics behave as fully five-dimensional black holes in the vicinity of the horizon, 
while they behave as four-dimensional black holes in the region far away from the horizon. 
That is, 
the squashed Kaluza-Klein black hole solutions 
asymptote to locally flat spacetimes with a compactified extra dimension   
and we can regard a series of these solutions as one of realistic higher-dimensional black hole models.  
Several aspects of squashed Kaluza-Klein black holes are also discussed, for example, 
thermodynamics \cite{Cai:2006td,Kurita:2007hu,Kurita:2008mj}, 
Hawking radiation \cite{Ishihara:2007ni,Chen:2007pu,Wei:2009kg}, 
quasinormal modes \cite{Ishihara:2008re,He:2008im,He:2008kq}, 
stabilities \cite{Kimura:2007cr,Nishikawa:2010zg},  
geodetic precession \cite{Matsuno:2009nz} and 
strong gravitational lensing \cite{Liu:2010wh}.

In this paper, we consider the Hawking radiation 
from a five-dimensional charged static Kaluza-Klein black hole with squashed horizons  
on the basis of the tunneling mechanism.    
To the best our knowledge, 
derivations of Hawking radiation in the framework of the tunneling method 
have not been discussed in asymptotically Kaluza-Klein spacetimes.   
In the present work, 
we extend the derivation of Hawking radiation as a tunneling process 
by using the technique of the dimensional reduction near the horizon 
in four dimensions  
to the case of the five-dimensional squashed Kaluza-Klein black hole.    
For simplicity, 
we restrict ourselves to uncharged radiations coming from the black hole 
and take into account the back reaction of the radiation.   
Then, in contrast to the previous studies \cite{Ishihara:2007ni,Chen:2007pu,Wei:2009kg},  
we obtain not only the Hawking temperature of the black hole but also 
the effect of the back reaction associated with the radiation in a very simple manner.

This paper is organized as follows. 
In Section \ref{review}, we review the properties of 
five-dimensional charged static Kaluza-Klein black hole solutions with squashed horizons. 
In Section \ref{reduction}, 
we show that the five-dimensional squashed Kaluza-Klein black hole metric effectively 
becomes a two-dimensional spherically symmetric metric 
by using the technique of the dimensional reduction near the horizon. 
In Section \ref{hawkingradiation}, 
we apply the tunneling mechanism to the squashed Kaluza-Klein black hole and 
derive both the Hawking temperature and the effect of the back reaction associated with the radiation.    
Section \ref{discussion} is devoted to discussion and conclusion.

\section{Review of squashed Kaluza-Klein black holes}\label{review}

We consider the charged static Kaluza-Klein black hole with squashed $S^3$ horizons, 
which is one of the exact solutions of 
the five-dimensional Einstein-Maxwell theory \cite{Ishihara:2005dp}. 
The metric and the gauge potential are written as  
\begin{eqnarray}
 ds^2 &=& - F(\rho ) dt ^2 + \frac{K^2 (\rho )}{F(\rho )} d\rho ^2 + \rho ^2 K^2 (\rho ) d\Omega _{S^2} ^2 
 + \frac{r_\infty ^2}{4 K^2 (\rho )} \left( d\psi + \cos \theta d\phi \right) ^2 ,  
\label{met} 
\\
 A_\mu &=& \left( \pm \frac{\sqrt{3 \rho _+  \rho _-}}{2 \rho }, 0, 0, 0, 0 \right) ,   
\label{gau}
\end{eqnarray}
where $d\Omega _{S^2} ^2 = d\theta ^2 + \sin ^2 \theta d\phi ^2$ 
denotes the metric of the unit two-sphere, and  
the functions $F(\rho )$ and $K(\rho )$ are given by  
\begin{eqnarray}\label{funcs}
 F(\rho ) = \frac{\rho ^2 - 2 M \rho + Q^2}{\rho ^2} 
= \frac{\left( \rho - \rho _+ \right) \left( \rho - \rho _- \right)}{\rho ^2} , 
\quad 
K^2 (\rho ) = \frac{\rho + \rho _0}{\rho } .
\end{eqnarray}
The parameters $r_\infty ,~ \rho _\pm$, and $\rho _0$ are related as 
$r_\infty ^2 = 4 \left( \rho _+ + \rho _0 \right) \left( \rho _- + \rho _0 \right)$, and 
the parameters $M$ and $Q$ denote the Komar mass and the charge of the black hole \eqref{met}, 
respectively. 
The coordinates $(t , \rho , \theta , \phi , \psi )$ run the ranges of 
$-\infty < t < \infty ,~ 0 < \rho < \infty ,~ 0 \leq \theta \leq \pi ,~ 
0 \leq \phi \leq 2 \pi$, and $0 \leq \psi \leq 4 \pi$, respectively.  
To avoid the existence of naked singularities on and outside the horizon, 
we choose the parameters such that 
$\rho _+ \geq \rho _- \geq 0$ and $\rho _- + \rho _0 > 0$.

The outer and the inner horizons are located at 
$\rho = \rho _+ = M + \sqrt{M^2 - Q^2} $ and 
$\rho = \rho _- = M - \sqrt{M^2 - Q^2} $, respectively.  
The shapes of horizons are the squashed $S^3$ in the form of the Hopf bundle.   
The surface gravity on the outer horizon of the black hole \eqref{met} is obtained as 
\begin{eqnarray}\label{surfaceg}
 \kappa 
= \frac{\rho _+ - \rho _-}{2 \rho _+ ^2} \sqrt\frac{\rho _+}{\rho _+ + \rho _0} . 
\end{eqnarray}
At infinity, $\rho \to \infty$, the metric \eqref{met} behaves as
\begin{eqnarray}
 ds^2 \simeq - dt ^2 + d\rho ^2 + \rho ^2 d\Omega _{S^2} ^2 
 + \frac{r_\infty ^2}{4} \left( d\psi + \cos \theta d\phi \right) ^2 . 
\end{eqnarray}
That is, the metric \eqref{met} asymptotes to a twisted constant $S^1$
fiber bundle over the four-dimensional Minkowski spacetime.
The size of the compactified extra dimension of the spacetime \eqref{met} 
at infinity is given by $r_\infty$.

Here, we discuss the physical meanings of the parameter $\rho _0$ in the metric \eqref{met}. 
If $\rho _0 \ll \rho _\pm$, 
$\rho $ dependence of the function $F(\rho )$ is important 
but the function $K(\rho )$ is almost unity for an observer outside the horizons, 
$\rho _0 \ll \rho _\pm \lesssim \rho $. 
Then, the observer feels the spacetime as 
the four-dimensional Reissner-Nordstr\"om black hole with a twisted constant $S^1$ fiber
with the metric 
\begin{eqnarray}\label{limit1}
 ds^2 \simeq 
    - \frac{\left( \rho - \rho _+ \right) \left( \rho - \rho _- \right)}{\rho ^2} dt ^2 
	+ \frac{\rho ^2}{\left( \rho - \rho _+ \right) \left( \rho - \rho _- \right)} d\rho ^2 
	+ \rho ^2 d \Omega_{S^2} ^2 
    + \frac{r_\infty^2}{4} \left( d \psi + \cos \theta d \phi \right) ^2 . 
\end{eqnarray}
On the other hand, 
if $\rho _\pm \ll \rho _0$, 
the function $K(\rho )$ becomes important for an observer at 
$\rho _\pm \lesssim \rho \ll \rho _0 $. 
With the help of a new coordinate $r = 2 \sqrt{\rho _0 \rho }$
and parameters $r_\pm = 2 \sqrt{\rho _0 \rho _\pm}$, 
since $r_\pm ^2 \ll r_\infty ^2$, 
the metric \eqref{met} approaches 
the five-dimensional Reissner-Nordstr\"om black hole \cite{Tangherlini:1963bw}:  
\begin{eqnarray}\label{limit2}
 ds^2 \simeq 
       - \frac{\left( r^2 - r_+^2 \right) \left( r^2 - r_-^2 \right)}{r ^4} dt^2 
       + \frac{r ^4}{\left( r^2 - r_+^2 \right) \left( r^2 - r_-^2 \right)} dr^2 
       + r^2 d \Omega_{S^3} ^2,    
\end{eqnarray}
where $d\Omega_{S^3}^2$ denotes the metric of the unit three-sphere.
Then, the observer feels the spacetime as an almost $S^3$ symmetric black hole. 
Therefore, the parameter $\rho _0$ gives 
the typical scale of transition from five dimensions to effective four dimensions.

\section{Dimensional reduction near the horizon}\label{reduction}

In this section, we discuss the technique of the dimensional reduction near the horizon 
in the squashed Kaluza-Klein black hole. 
Considering the near horizon behavior of the action for a five-dimensional complex scalar field 
in the squashed Kaluza-Klein black hole background \eqref{met}, 
we show that the five-dimensional action is reduced to a two-dimensional one and 
the five-dimensional squashed Kaluza-Klein black hole behaves as 
an effectively two-dimensional black hole in the region very close to the horizon.

We consider the five-dimensional complex scalar field $\Phi$ 
in the squashed Kaluza-Klein black hole background \eqref{met}. 
The action is given by 
\begin{eqnarray}\label{action0}
S = \int d^5 x \sqrt{-g} g^{\mu \nu } \left( \partial _\mu + i e A_\mu  \right)
\Phi ^* \left( \partial _\nu - i e A_\nu  \right) \Phi 
+ S_\text{int} ,
\end{eqnarray}
where the first term is the kinetic term and 
the second term $S_\text{int}$ represents the mass, the potential and the interaction terms. 
Substituting the metric \eqref{met} and the gauge potential \eqref{gau} 
into the action \eqref{action0}, we have 
\begin{eqnarray}\label{action01}
S &=&- \frac{r_\infty}{2} \int dt d\rho d\theta d\phi d\psi 
\sin \theta \rho ^2 K^2 \Phi ^*  
\left[ 
- \frac{1}{F}
\left( \partial _t - i e A_t \right) ^2 +
\frac{1}{\rho ^2 K^2} \partial _\rho \left( \rho ^2 F \partial _\rho \right) 
\right.
\notag \\
& &+ \left. \frac{1}{\rho ^2 K^2 \sin \theta } \partial _\theta \left( \sin \theta \partial _\theta \right) +
\frac{1}{\rho ^2 K^2} \left( \frac{1}{ \sin \theta } \partial _\phi 
- \frac{\cos \theta}{ \sin \theta } \partial _\psi  \right)^2 +
\frac{4 K^2}{r_\infty ^2} \partial _\psi ^2
\right] \Phi 
+ S_\text{int} .  
\end{eqnarray}
We perform the partial wave decomposition of $\Phi$ as   
\begin{eqnarray}
 \Phi = \sum _{l, m, \lambda } \Phi _{l m \lambda }(t, \rho ) S_{l m \lambda } (\theta ) 
e^{i(m\phi + \lambda \psi )} , 
\end{eqnarray}
where the function $S_{l m \lambda }  $ is the spin-weighted spherical function, which satisfies 
\begin{eqnarray}
 \left[ 
\frac{1}{\sin \theta } \partial _\theta \left( \sin \theta \partial _\theta \right) 
- \frac{\left( m - \lambda \cos \theta \right)^2}{\sin ^2 \theta } + l(l+1)-\lambda ^2 
\right] S_{l m \lambda } = 0 . 
\end{eqnarray}
We have the conditions such that 
$l \geq \lambda$ and $m = -l, -l+1, \cdots, l$. 
The periodicity also requires that 
$2 \lambda , 2 m $, and $\lambda \pm m$ are integers. 
Then we obtain the action \eqref{action01} in the form   
\begin{eqnarray}\label{action1}
S &=&- \frac{r_\infty}{2} \int dt d\rho d\theta d\phi d\psi 
\sin \theta \rho ^2 K^2 \sum _{l', m', \lambda' } \Phi _{l' m' \lambda' } ^* S_{l' m' \lambda' } ^*
e^{-i(m'\phi + \lambda' \psi )}
\notag \\
& &\times
\left[ 
- \frac{1}{F}
\left( \partial _t - i e A_t \right) ^2 +
\frac{1}{\rho ^2 K^2} \partial _\rho \left( \rho ^2 F \partial _\rho \right) - 
\frac{l(l+1)-\lambda ^2}{\rho ^2 K^2} - \frac{4 K^2 \lambda ^2}{r_\infty ^2} 
\right] 
\notag \\
& &\times 
\sum _{l, m, \lambda } \Phi _{l m \lambda } S_{l m \lambda }  
e^{i(m\phi + \lambda \psi )} 
+ S_\text{int} . 
\end{eqnarray}
Define the tortoise coordinate as 
\begin{eqnarray}
d\rho _* = \frac{K}{F} d\rho ,
\end{eqnarray}
the action \eqref{action1} is rewritten as 
\begin{eqnarray}\label{action2} 
S &=&- \frac{r_\infty}{2} \int dt d\rho _* d\theta d\phi d\psi 
\sin \theta \rho ^2 K \sum _{l', m', \lambda' } \Phi _{l' m' \lambda' } ^* S_{l' m' \lambda' } ^*
e^{-i(m'\phi + \lambda' \psi )}
\notag \\
& &\times
\left[ 
- \left( \partial _t - i e A_t \right) ^2 +
\frac{1}{\rho ^2 K} \partial _{\rho_*} \left( \rho ^2 K \partial _{\rho_*} \right) - 
F \left\{ \frac{l(l+1)-\lambda ^2}{\rho ^2 K^2} + \frac{4 K^2 \lambda ^2}{r_\infty ^2} \right\}
\right]  
\notag \\
& &\times 
\sum _{l, m, \lambda } \Phi _{l m \lambda } S_{l m \lambda }  
e^{i(m\phi + \lambda \psi )} 
+ S_\text{int} . 
\end{eqnarray}

We consider the action \eqref{action2} in the region near the black hole horizon, $\rho \simeq \rho _+$. 
Since the theory becomes the high-energy theory near the horizon and 
$F(\rho _+) = 0 $ at $\rho \to \rho _+$,  
we ignore all terms in $S_\text{int}$ and 
retain the dominant kinetic term only.  
Then we have   
\begin{eqnarray}\label{action3} 
S &=&- \frac{r_\infty}{2} \int dt d\rho _* d\theta d\phi d\psi 
\sin \theta \rho ^2 K \sum _{l', m', \lambda' } \Phi _{l' m' \lambda' } ^* S_{l' m' \lambda' } ^*
e^{-i(m'\phi + \lambda' \psi )}
\notag \\
& &\times
\left[ 
- \left( \partial _t - i e A_t \right) ^2 +
\frac{1}{\rho ^2 K} \partial _{\rho_*} \left( \rho ^2 K \partial _{\rho_*} \right) 
\right]  
\sum _{l, m, \lambda } \Phi _{l m \lambda } S_{l m \lambda }  
e^{i(m\phi + \lambda \psi )} .
\end{eqnarray}
We return to the expression written in terms of the coordinate $\rho$, 
the action \eqref{action3} takes the form  
\begin{eqnarray}\label{action4} 
S = - 4 \pi ^2 r_\infty \sum _{l, m, \lambda } \int dt d\rho
\rho ^2 K \Phi _{l m \lambda } ^* 
\left[ 
- \frac{K}{F} \left( \partial _t - i e A_t \right) ^2 + 
\partial _{\rho} \left( \frac{F}{K} \partial _{\rho} \right) 
\right]  
\Phi _{l m \lambda } ,
\end{eqnarray}
where we have used the orthonormal condition for the spin-weighted spherical function, 
\begin{eqnarray}
 \int d\theta \sin \theta S_{l' m' \lambda' } ^* S_{l m \lambda } 
= \delta _{l' , l} \delta _{m' , m} \delta _{\lambda' , \lambda}  ,
\end{eqnarray}
and integrated the angular coordinate parts. 
From the action \eqref{action4}, we can regard  
the field $\Phi _{l m \lambda }$ as a two-dimensional complex scalar field 
in the backgrounds with the metric $ds^2_\text{2D}$, 
the gauge potential $A_i$ $(i = t , \rho )$, and the dilaton field $\Psi$: 
\begin{eqnarray}
 & & 
 ds^2_\text{2D} = - \frac{F}{K} dt ^2 + \frac{K}{F} d\rho ^2 , \label{effmet} \\
& & 
 A_t = \pm \frac{\sqrt{3 \rho _+ \rho _-}}{2 \rho } , \quad 
 A_\rho = 0 ,  
\\
& & 
 \Psi = \rho ^2 K ,  
\end{eqnarray}
where the functions $F$ and $K$ are given by the equations \eqref{funcs}.

We see that the metric of the five-dimensional squashed Kaluza-Klein black hole \eqref{met} 
behaves as an effectively two-dimensional 
metric \eqref{effmet} 
by using the technique of the dimensional reduction near the horizon. 
For confirmation, 
we calculate the surface gravity on the horizon 
of the two-dimensional spacetime \eqref{effmet} as  
\begin{eqnarray}\label{redsurfaceg}
 \kappa = \frac{1}{2} \partial _\rho \left. \left( \frac{F}{K} \right) \right|_{\rho = \rho _+} 
= \frac{\rho _+ - \rho _-}{2 \rho _+ ^2} \sqrt\frac{\rho _+}{\rho _+ + \rho _0}  .
\end{eqnarray}
We see that 
this surface gravity \eqref{redsurfaceg} coincides with that of 
the five-dimensional squashed Kaluza-Klein black hole \eqref{surfaceg}.

\section{Hawking radiation as tunneling process}\label{hawkingradiation}


In this section, 
we derive both the Hawking temperature and 
the effect of the back reaction associated with the radiation 
in the squashed Kaluza-Klein black hole background 
on the basis of the tunneling mechanism.  
Here we use the two-dimensional effective metric \eqref{effmet} as the background metric. 
Since the tunneling effect is a quantum one arising within the Planck
length near the horizon region,
our derivation of Hawking radiation as a tunneling process 
by using the technique of the dimensional reduction is justified. 

To describe the across-horizon phenomena, 
Parikh and Wilczek used the Painlev\'e coordinates, 
where the coordinate singularity at the horizon was removed \cite{Parikh:1999mf}. 
Similarly, 
we define the Painlev\'e-like coordinate $t_p$ to the metric \eqref{effmet} by  
\begin{eqnarray}
 dt_p &=& dt + \frac{K}{F} \sqrt{1 - \frac{F}{K}} d\rho .  
\end{eqnarray}
Then the metric \eqref{effmet} is rewritten as  
\begin{eqnarray}\label{effmet2}
 ds^2_\text{2D} 
= - \frac{\rho ^2 - 2 M \rho + Q^2}{\rho ^2 K} dt_p ^2 
+ 2 \sqrt{\frac{K - 1}{K} + \frac{2 M}{\rho K} - \frac{Q^2}{\rho ^2 K}} dt_p d\rho + d\rho ^2 . 
\end{eqnarray}
The metric \eqref{effmet2} has a number of interesting features.  
At any fixed time, $t_p = \text{const.}$, the spatial geometry is flat, 
and for any fixed radius, $\rho = \text{const.}$, 
the boundary geometry is the same as that of the metric \eqref{effmet}.


The radial null geodesics for the metric \eqref{effmet2} are given by $ds^2_\text{2D} = 0$. 
We obtain  
\begin{eqnarray}\label{nullgeo}
 \dot \rho = \frac{d\rho }{dt_p} 
= \pm 1 - \sqrt{\frac{K - 1}{K} + \frac{2 M}{\rho K} - \frac{Q^2}{\rho ^2 K}}  , 
\end{eqnarray}
where the upper and the lower signs correspond to the outgoing and the ingoing geodesics, 
respectively, under the implicit assumption that $t_p$ increases towards the future.  
Since we can ignore the mass of matter field near the horizon region by the discussion of dimensional reduction, 
the use of null geodesics \eqref{nullgeo} in our derivation of Hawking radiation is justified. 
If we restrict ourselves to uncharged radiations coming from the black hole 
and take into account the effect of the particle's self gravitation, i.e., 
the back reaction of the radiation,   
we replace the mass of the black hole $M$ by $M - \omega$, 
where $\omega$ denotes the energy of the particle  
which escapes from the black hole by the tunneling mechanism.   
Then we have the metric \eqref{effmet2} and the radial null geodesics \eqref{nullgeo} in the forms 
\begin{eqnarray}
 ds^2_\text{2D} &=& 
- \frac{\rho ^2 - 2 (M - \omega) \rho + Q^2}{\rho ^2 K} dt_p ^2 
+ 2 \sqrt{\frac{K - 1}{K} + \frac{2 (M - \omega)}{\rho K} - \frac{Q^2}{\rho ^2 K}} dt_p d\rho + d\rho ^2 , 
\label{effmet3} \\ 
 \dot \rho &=& \pm 1 - \sqrt{\frac{K - 1}{K} + \frac{2 (M - \omega )}{\rho K} - \frac{Q^2}{\rho ^2 K}} . 
\label{nullgeo2}
\end{eqnarray}


By following the method proposed by Parikh and Wilczek, we evaluate the WKB probability
amplitude for a classically forbidden trajectory. The imaginary
part of the action for an outgoing positive energy particle,
which crosses the horizon outwards 
from $\rho = \rho_\text{in}$ to $\rho = \rho_\text{out}$, is given by
\begin{eqnarray}
 \text{Im} S = \text{Im} \int^{\rho_\text{out}}_{\rho_\text{in}} p_\rho d\rho 
 = \text{Im} \int^{\rho_\text{out}}_{\rho_\text{in}} \int^{p_\rho}_{0} d p_\rho ' d\rho 
 = \text{Im} \int^{\rho_\text{out}}_{\rho_\text{in}} \int^{H}_{0} \frac{dH'}{\dot \rho } d \rho ,
\end{eqnarray}
where we have multiplied and divided the integrand by the two sides of Hamilton's equation,
$\dot \rho = dH / dp_\rho |_\rho $,  
to change the variable from the momentum to the energy.  
Taking into account the effect of the back reaction of the radiation, we obtain  
\begin{eqnarray}
 \text{Im} S 
= \text{Im} \int^{M - \omega }_{M} \int^{\rho_\text{out}}_{\rho_\text{in}} \frac{d\rho}{\dot \rho } dH'
= \text{Im} \int^{ \omega }_{0} \int^{\rho_\text{out}}_{\rho_\text{in}} 
\frac{d\rho}{1 - \sqrt{\frac{K - 1}{K} + \frac{2 (M - \omega ')}{\rho K} - \frac{Q^2}{\rho ^2 K}} } (-d\omega ') , 
\end{eqnarray}
where we have used the Hamiltonian, $H = M - \omega $, and 
the radial null geodesic \eqref{nullgeo2} with the upper sign.    
Now the integral can be done by deforming the contour, 
so as to ensure that positive energy solutions decay in time, 
that is, into the lower half $\omega '$ plane.  
By using Feynman's $i \epsilon $ prescription, $\omega \to \omega - i \epsilon $, 
we have 
\begin{eqnarray}
 \text{Im} S &=& 
     \text{Im} \int^{\rho_\text{out}}_{\rho_\text{in}}  \int^{M - \omega }_{M} 
\frac{dM'}{1 - \sqrt{\frac{K - 1}{K} + \frac{2 M'}{\rho K} - \frac{Q^2}{\rho ^2 K}} } d\rho 
\notag \\
 &=& \text{Im} \int^{\rho_\text{out}}_{\rho_\text{in}}  \int^{M - \omega }_{M} 
\frac{dM'}{1 - \sqrt{\frac{K - 1}{K} + \frac{2 M'}{\rho K} - \frac{Q^2}{\rho ^2 K}} + i \epsilon } d\rho 
\notag \\
 &=& \text{Im} \int^{\rho_\text{out}}_{\rho_\text{in}}  \int^{M - \omega }_{M} 
- i \pi \delta \left( 1 - \sqrt{\frac{K - 1}{K} + \frac{2 M'}{\rho K} - \frac{Q^2}{\rho ^2 K}} \right) dM' d\rho 
\notag \\
 &=& - \pi \int^{\rho_\text{out}}_{\rho_\text{in}} \rho K d\rho .
\label{sekibun}
\end{eqnarray}

Thus we obtain the WKB probability amplitude as     
\begin{eqnarray}\label{wkb}
 \Gamma &\simeq& e^{-2 \text{Im} S} 
\notag \\
 &=& \exp \left[ 
\frac{\pi}{2}  
\left( 2 \rho_\text{out} + \rho _0 \right) 
\sqrt{\rho_\text{out} \left(\rho_\text{out} + \rho _0 \right)}
- \frac{\pi}{2} \left( 2 \rho_\text{in} + \rho _0 \right) 
\sqrt{\rho_\text{in} \left(\rho_\text{in} + \rho _0 \right)} \right.
\notag \\
& & \left. - \frac{\pi \rho _0 ^2}{2} \log 
\frac{\sqrt{\rho_\text{out}} + \sqrt{\rho_\text{out} + \rho _0}}
{\sqrt{\rho_\text{in}} + \sqrt{\rho_\text{in} + \rho _0}}
\right] ,
\end{eqnarray}
where the explicit forms of $\rho_\text{in}$ and $\rho_\text{out}$ are, respectively, given by  
\begin{eqnarray}\label{bound}
\rho_\text{in} = \rho _+ = M + \sqrt{M^2 - Q^2} , \quad 
\rho_\text{out} = M - \omega + \sqrt{(M - \omega )^2 - Q^2} . 
\end{eqnarray}

If we consider the effect in the second order of $\omega $, 
in terms of $\rho _+$ and $\rho _-$, 
the WKB probability amplitude \eqref{wkb} takes the form  
\begin{eqnarray}\label{wkbfinal}
 \Gamma 
 \simeq 
\exp \left( 
- \frac{2 \pi}{\tilde \kappa} \omega 
+ \alpha \omega ^2 
\right)  ,
\end{eqnarray}
where the constants $\tilde \kappa$ and $\alpha $ are, respectively, given by    
\begin{eqnarray}
 \tilde \kappa &=& \frac{\rho _+ - \rho _-}{2 \rho _+ ^2} \sqrt\frac{\rho _+}{\rho _+ + \rho _0}  ,
\label{surfacegder} \\
 \alpha &=& \frac{2 \pi \rho _+ 
\left[ 2 \rho _+ (\rho _+ - 3 \rho _-) + \rho _0 (\rho _+ - 5 \rho _-) \right]}{( \rho _+ - \rho _- )^3}
\sqrt\frac{\rho _+}{\rho _+ + \rho _0} .
\label{backreaction}
\end{eqnarray}
We see that the constant $\tilde \kappa$ \eqref{surfacegder} coincides with 
the surface gravity of the black hole \eqref{surfaceg}, that is, 
$\tilde \kappa = \kappa $. 
Thus, by comparing the WKB probability amplitude \eqref{wkbfinal} to the first order in $\omega$ 
with the Boltzmann factor in a thermal equilibrium state at the temperature $T$, 
$\Gamma = e ^{-\omega / T}$,  
we obtain the Hawking temperature of the five-dimensional squashed Kaluza-Klein black hole \eqref{met} as  
\begin{eqnarray}
 T_\text{H} = \frac{\tilde \kappa}{2 \pi} 
= \frac{\rho _+ - \rho _-}{4 \pi \rho _+ ^2} \sqrt\frac{\rho _+}{\rho _+ + \rho _0} ,
\end{eqnarray}
which is the desired result \cite{Cai:2006td,Kurita:2007hu,Kurita:2008mj,Ishihara:2007ni,Wei:2009kg},    
and we can regard the quadratic term of $\omega$ in the equation \eqref{wkbfinal}, $\alpha \omega^2$, 
as the correction by the back reaction of the radiation.

In the limit, $\rho _0 \to 0$, 
we obtain the metric \eqref{met} with $K = 1$, 
which locally has the geometry of the black string \eqref{limit1}. 
In this limit, 
the WKB probability amplitude \eqref{wkb} 
reduces to 
\begin{eqnarray}\label{wkb4d}
 \Gamma &\simeq& 
\exp \left[ 
\pi
\left( \rho_\text{out}^2 - \rho_\text{in}^2 \right)
\right] 
\notag \\
&=& 
\exp \left[ 
- 4 \pi \omega \left( M - \frac{\omega }{2} \right) 
+ 2 \pi \left( M -\omega \right) \sqrt{(M - \omega )^2 - Q^2} -2 \pi M \sqrt{M^2 - Q^2} 
\right] .
\end{eqnarray}
To the first order in $\omega$, 
this WKB probability amplitude \eqref{wkb4d} is consistent with previous results 
of thermal emission at the Hawking temperature $T_\text{H}$ for 
the four-dimensional Reissner-Nordstr\"om black hole \cite{Hawking:1974sw,Parikh:1999mf}:   
\begin{eqnarray}
 T_\text{H} = \frac{\sqrt{M^2 - Q^2}}{2 \pi \left( M + \sqrt{M^2 - Q^2} \right) ^2} . 
\end{eqnarray}

\section{Discussion and Conclusion}\label{discussion}

To sum up,   
we have extended the derivation of Hawking radiation 
by using the tunneling mechanism 
in four-dimensional black hole backgrounds 
to the case of the five-dimensional charged static Kaluza-Klein black hole with squashed horizons.  
We have restricted ourselves to uncharged radiations coming from the black hole 
and taken into account the back reaction of the radiation.   
Then 
we have derived the Hawking temperature of the black hole 
in a very simple manner.   
The calculation in our derivation 
by using the technique of the dimensional reduction near the horizon 
is little different from the original calculation for 
four-dimensional spherically symmetric black holes   
discussed in \cite{Parikh:1999mf}. 
Although, our method has included the essential effect of the extra dimension,   
we have derived the Hawking temperature correctly by evaluating the effect to the first order in 
the energy of the particle which escapes from the black hole by the tunneling mechanism. 
We have seen that 
the technique of the dimensional reduction has been useful in the derivation of Hawking radiation 
on the basis of the tunneling mechanism    
in the five-dimensional squashed Kaluza-Klein black hole backgrounds.   
In contrast to the previous studies of Hawking radiation from squashed Kaluza-Klein black holes 
\cite{Ishihara:2007ni,Chen:2007pu,Wei:2009kg}, 
we have derived not only the Hawking temperature but also 
the effect of the back reaction associated with the radiation.  
If higher-dimensional black holes are created in future accelerator experiments 
and we assume that the five-dimensional squashed Kaluza-Klein black hole solutions describe 
geometries around such black holes,     
we expect that our present work could make a contribution to the verifications of 
Hawking radiation and extra dimensions   
in asymptotically Kaluza-Klein spacetimes.

In this paper, 
our analysis by using the tunneling mechanism has been confined only to 
the derivation of the Hawking temperature of the five-dimensional squashed Kaluza-Klein black hole   
and 
we have not discussed the blackbody spectrum.  
Then there remains the possibility that 
the black hole is not the blackbody but merely the thermal body. 
However, we expect that 
we can obtain the blackbody spectrum of the Hawking radiation 
in the squashed Kaluza-Klein black hole background  
by applying the methods in \cite{Banerjee:2009wb,Umetsu:2009ra,Umetsu:2010ts}, 
where the derivation of the blackbody spectrum with the Hawking temperature from
the expectation value of number operator 
by using the properties of the tunneling mechanism is discussed.      
We leave the analysis of the blackbody spectrum for the five-dimensional squashed Kaluza-Klein black hole 
on the basis of the tunneling mechanism for the future.


\end{document}